\documentclass[apj]{emulateapj} 

\newcommand{\beq}{\begin{equation}}
\newcommand{\eeq}{\end{equation}}
\newcommand{\lta}{\la}

\newcommand{\msunh}{\>h^{-1}\rm M_\odot}
\newcommand{\mpch}{\>h^{-1}\rm Mpc}
\newcommand{\kpch}{\>h^{-1}\rm kpc}
\newcommand{\etal}{{et al.}}

\shorttitle{The Accretion of Dark Matter Subhalos}
\shortauthors{Yang et al.}

\begin{document}


\title{An analytical model for the accretion of dark matter subhalos}

\author{Xiaohu Yang\altaffilmark{1}, H.J. Mo\altaffilmark{2}, Youcai
  Zhang\altaffilmark{1,4}, Frank C. van den Bosch\altaffilmark{3} }

\altaffiltext{1}{ Key Laboratory for Research in Galaxies and Cosmology, Shanghai
  Astronomical Observatory; the Partner Group of MPA; Nandan Road 80, Shanghai
  200030, China;  E-mail: xhyang@shao.ac.cn}

\altaffiltext{2}{Department of Astronomy, University of Massachusetts, Amherst
  MA 01003-9305}

\altaffiltext{3} {Astronomy Department, Yale University, P.O. Box 208101, New
  Haven, CT 06520-8101}

\altaffiltext{4}{Graduate School of the Chinese Academy of Sciences, 19A,
  Yuquan Road, Beijing, China}


\begin{abstract}
  An analytical model  is developed for the mass function  of cold dark matter
  subhalos  at  the  time of  accretion  and  for  the distribution  of  their
  accretion times.  Our model is based on  the model of Zhao et al. (2009) for
  the median assembly  histories of dark matter halos,  combined with a simple
  log-normal distribution to describe the scatter in the main-branch mass at a
  given time for halos of the same final mass. Our model is simple, and can be
  used to predict  the un-evolved subhalo mass function,  the mass function of
  subhalos  accreted  at a  given  time,  the  accretion-time distribution  of
  subhalos of  a given initial mass, and  the frequency of major  mergers as a
  function  of time.   We test  our model  using  high-resolution cosmological
  $N$-body  simulations,  and  find  that  our  model  predictions  match  the
  simulation results remarkably well.  Finally, we discuss the implications of
  our  model  for  the evolution  of  subhalos  in  their  hosts and  for  the
  construction of  a self-consistent  model to link  galaxies and  dark matter
  halos at different cosmic times.
\end{abstract}




\section{Introduction}

In  the  current  Cold  Dark  Matter (hereafter  CDM)  paradigm  of  structure
formation, a key  concept in the build-up of structure in  the universe is the
hierarchical  formation of  dark matter  halos.  Galaxies  and  other luminous
objects  are assumed to  form by  cooling and  condensation of  baryons within
these halos \citep[see][for an overview]{MBW10}.  In this scenario, a detailed
understanding  of the  formation  and structure  of  dark matter  halos is  of
fundamental importance for predicting the properties of luminous objects, such
as galaxies and clusters of galaxies.
 
The formation history of a CDM  halo is conveniently represented by its merger
tree, which describes how its  progenitors merge and accrete during its entire
formation history.  For  a given cosmological model, such  merger trees can be
constructed either  from N-body  simulations or from  Monte-Carlo realizations
based on the extended  Press-Schechter (PS) formalism \citep[][]{PS74, BCEK91,
  Bower91,  LC93,SMT01}.   In recent  years,  much  effort  has been  made  to
characterize and understand the statistical  properties of halo formation in a
CDM cosmogony.   One particular  aspect of the  halo formation process  is the
existence  of a subhalo  population, which  is produced  by the  accretion and
survival  of progenitors  at various  times  \citep[e.g.][]{Klypin99, Moore99,
  Kravtsov04,  Gao04,  Del04, vdB05,  DKM07,  Gioc08a,  Gioc08b, Sp08,  Wet09,
  Angulo09,  LM09}.   Since  galaxies  may   form  at  the  centers  of  these
progenitors  and   merge  into   the  final  halo   along  with   their  hosts
\citep[e.g.][]{Kang05}, the  statistical properties of  the subhalo population
are expected to  be closely linked to those of satellite  galaxies. One of the
basic  properties  of the  subhalo  population is  the  mass  function of  the
progenitors of subhalos  (i.e., the masses of the subhalos  at their moment of
accretion). Following  \citet{vdB05}, we  refer to this  mass function  as the
{\it  un-evolved} subhalo  mass  function,  to distinguish  it  from the  {\it
  evolved} subhalo mass function that refers to the present day masses of dark
matter  subhalos  (see \S\ref{sec_model}  for  details).  A  number of  recent
investigations  have  used these  subhalo  mass  functions  in an  attempt  to
characterize   the   galaxy-dark  matter   connection   across  cosmic   times
\citep[e.g.][]{   Vale04,  Vale06,   Zheng05,  Zheng07,   Conroy06,  Conroy07,
  Conroy09, Yang09, Yang11, Li09, Moster10, BCW10, WJ10, Wetzel10, Neistein11,
  AvilaReese11}.

So far  the subhalo mass function  has been studied  with $N$-body simulations
and  Monte-Carlo  realizations of  the  extended  PS formalism  \citep[e.g.][]
{Sheth99, Somerville99,  Cole00, vdB05, Gioc08a, Cole08,  PCH08, FM08, FMB10}.
In this  paper we show  that a simple  analytical model can be  constructed to
describe not only the mass  distribution of subhalos but also the distribution
of their accretion times. We  use $N$-body simulations to demonstrate that the
model is remarkably  accurate. The model is not only  simple to implement, but
also  provides important  insights into  the formation  and evolution  of dark
matter   halos  and  subhalos.    Furthermore,  as   we  briefly   discuss  in
\S\ref{sec_discussion}, our model also  provides a self-consistent way to link
galaxies and dark matter halos at different redshifts.

The structure of the paper is organized as follows. In Section \ref{sec_Nbody}
we outline the simulations based on  which we will test our model.  In Section
\ref{sec_model} we  describe our  model. In Section~\ref{sec_test}  we present
our model predictions for the conditional mass function of subhalos, the major
merger  rates, and  the subhalo  mass  function. These  model predictions  are
tested  against   $N$-body  simulation  results   in  Section  \ref{sec_test}.
Finally, in  Section~\ref{sec_discussion} we  discuss the universality  of our
model for other models of structure formation, and discus how our model can be
used  to study  the evolution  of  subhalos in  their hosts  and to  construct
self-consistent models that link galaxies  and dark matter halos across cosmic
times.

Throughout this paper, we use `$\ln$' to denote natural logarithm and `$\log$'
to denote the 10-based logarithm.

\section{The Simulations}
\label{sec_Nbody}

Before presenting our model, let us first describe briefly the
the $N$-body simulations to be used to test the model.

We use two  different $N$-body simulations that assume  the same cosmology but
use different box sizes (mass resolutions).  Both simulations were carried out
using  the  massively  parallel  GADGET2 code  \citep[][]{Sp01a,  Sp05}.   The
simulations  evolved  $1024^3$ dark  matter  particles  in  periodic boxes  of
$100\mpch$ and  $300\mpch$ on a  side, respectively, from redshift  $z=100$ to
the  present epoch  ($z=0$). The  particle masses  and softening  lengths are,
respectively,  $6.93\times10^7\msunh$ and $2.25\kpch$  for the  $100\mpch$ box
simulation, and $1.87\times10^9\msunh$ and  $6.75\kpch$ for the $300\mpch$ box
simulation.  The cosmological parameters used  in the simulations are based on
those  published   in  \citet[][]{Dunkley09}:  $\Omega_{\rm   m}=  \Omega_{\rm
  dm}+\Omega_{\rm     b}=0.258$,     $\Omega_{\rm    b}=0.044$,     $h=0.719$,
$\Omega_\Lambda=0.742$,   $n=0.963$,  and  $\sigma_8=0.796$.    Unless  stated
otherwise, our model predictions are  also for the $\Lambda$CDM cosmology with
this  particular set  of  parameters. For  both  simulations, a  total of  100
outputs in  equal $\log(1+z)$  interval were made,  starting from  $z=50$ (for
$300\mpch$ box) and $z=20$ (for $100\mpch$ box) to $z=0$.

Dark matter  halos were identified from  the simulations at  each output using
the  standard friends-of-friends  (FOF) algorithm  \citep[][]{Davis85}  with a
linking length of $0.2$ times the mean interparticle separation.  Here we keep
all halos  with at least 20  particles.  Based on halos  at different outputs,
halo merger trees were constructed  \citep[see][]{LC93}.  A halo in an earlier
output is considered to be a progenitor  of the present halo if more than half
of its particles are  found in the present halo.  The main  branch of a merger
tree is  defined to  consist of all  the progenitors  one goes through  as one
climbs from the bottom to the  top, choosing always the most massive branch at
every branching point.   These progenitors are referred to  as the main-branch
progenitors, and the time dependence of the main branch mass is referred to as
the assembly history. In our  analysis based on the $300\mpch$ box simulation,
we randomly choose  about 200 trees, from the total  merger tree catalogue, to
sample each of the two  massive bins at $M_h\sim 10^{14.5}\msunh$ and $M_h\sim
10^{14.0}\msunh$, and about 2000 trees to  sample each of the low-mass bins at
$M_h\sim  10^{13.0}\msunh$  and  $M_h\sim  10^{12.0}\msunh$, with  bin  widths
$\Delta  \log M_h\sim  0.7,0.1,0.1,0.02$, respectively.   Our tests  later are
mainly based on the $300\mpch$ box simulation.  However, in many cases we also
use halo merger  trees obtained from the $100\mpch$  box simulation to achieve
better  mass resolution. Specifically,  about 7,  36, 400  and 2000  trees are
selected from  this simulation to sample  the merger histories  for halos with
$M_h\sim 10^{14.5}\msunh$, $\sim  10^{14.0}\msunh$, $\sim 10^{13.0}\msunh$ and
$\sim 10^{12.0}\msunh$, with bin widths $\Delta \log M_h\sim 0.5,0.5,0.5,0.3$,
respectively.

\section{The Model}
\label{sec_model}

Now we  come back to our  modeling of the  accretion of subhalos.  We  want to
obtain the distribution of dark matter  subhalos with respect to their mass at
accretion, $m_a$, and their accretion redshift,  $z_a$, in a host halo of mass
$M_0$ at redshift $z_0$. For convenience we use %
\begin{equation}\label{eq:S_a0}
s_a\equiv \sigma_a^2=\sigma^2(m_a);
~~~~~~
S_0\equiv \sigma_0^2=\sigma^2(M_0)
\end{equation}
to label the masses, $m_a$ and $M_0$, and 
\begin{equation}\label{eq:delta_a0}
\delta_a\equiv \delta_c(z_a);
~~~~~~
\delta_0\equiv \delta_c(z_0)
\end{equation}
to label the redshifts $z_a$ and  $z_0$.  Here $\sigma (M)$ is the variance of
the linear density  field at $z=0$ on mass  scale $M$, and $\delta_c(z)\approx
1.686/D(z)$ [with $D(z)$ the linear  grow factor normalized to unity at $z=0$]
is the critical  density of spherical collapse at redshift  $z$.  We write the
mean number  of subhalos of  mass $m_a$ accreted  at redshift $z_a$ in  a host
halo $(M_0, \delta_0)$ as 
\begin{equation}\label{eq:d2N_a1}
{\rm d}^2N_a={\cal N}_a(s_a, \delta_a\vert S_0, \delta_0) {\rm d} \ln
m_a\,{\rm d} \ln (1+z_a)\,.
\end{equation}

The mean mass, ${\overline M}(z)$, of the main branch halos for all $(M_0,
\delta_0)$ - halos is in general a monotonically decreasing function of
redshift \citep[e.g.,][]{AvilaReese98,vdB02,Wec02}, and we write ${\overline
  M}_a\equiv {\overline M}(z_a)$. Hence, for given $(M_0, \delta_0)$ we can
use ${\overline M}_a$ as a time-variable. Denote by ${\cal F}(s_a,
\delta_a\vert S_0, \delta_0; {\overline M}_a)$ the mean fraction of the total
mass accreted in the `time-interval' $[{\overline M}_a - {\rm d} {\overline
    M}_a, {\overline M}_a]$ that is in halos of mass $m_a$.  We can write
\begin{equation}
{\rm d}^2 N_a ={1\over m_a}{\cal F}(s_a,\delta_a\vert S_0,\delta_0; 
{\overline M}_a) {\rm d} {\overline M}_a\, {\rm d} s_a\,.
\end{equation} 
It then follows that
\begin{equation}\label{eq:N_a_1}
{\cal N}_a
={\cal F}(s_a,\delta_a\vert S_0,\delta_0; {\overline M}_a)
{{\rm d} s_a\over {\rm d} m_a}
{{\rm d} {\overline M}_a\over {\rm d}\ln\delta_a}
{{\rm d}\ln\delta_a\over {\rm d}\ln(1+z_a)}\,.
\end{equation} 
Note that ${\rm d} s_a/{\rm d}m_a$ is determined by the perturbation power
spectrum, ${\rm d}\ln\delta_a/{\rm d}\ln(1+z_a)$ by the linear growth factor,
and ${\rm d} {\overline M} (z_a)/{\rm d}\ln\delta_a$ by the mean halo assembly
history.

We can integrate ${\cal N}_a$ over the mean mass assembly history to
obtain the so-called un-evolved subhalo mass function:
\begin{eqnarray}
{{\rm d} N_a\over {\rm d}\ln m_a}
&=&
\int{\cal N}_a (s_a, \delta_a\vert S_0, \delta_0) {\rm d}\ln (1+z_a)
\nonumber\\
&=&
\int_{m_a}^{M_0}
{\cal F} (s_a, \delta_a\vert S_0, \delta_0; {\overline M}_a) 
{{\rm d}s_a\over {\rm d} m_a}
{\rm d} {\overline M}_a\,.
\end{eqnarray}
This function describes the distribution of the masses at accretion of all
subhalos accreted into the main-branch of the merger history of the $(M_0,
\delta_0)$ host halo.  Thus for a given cosmology, one can obtain both ${\cal
  N}_a$ and the un-evolved subhalo mass function once models for ${\cal
  F}(s_a, \delta_a\vert S_0, \delta_0; {\overline M}_a)$ and for the mean halo
assembly history are adopted.  In general ${\cal F}$ can be obtained using
halo merger trees constructed either from numerical simulations or from
analytical models, such as the extended PS formalism. Here we develop a simple
analytical model based on the statistical properties of halo assembly
histories, instead of on the full merger trees.

Consider a halo whose main-branch mass is $M_a$ at $z_a$.  Since we are
modeling the accretion of subhalos into the main branch of its merger tree, we
must have that $m_a\le m_{\rm max}$ with\footnote{Note that $m_a$ reflects the
  mass of the progenitor halo which, after accretion into the main-branch,
  increases the main-branch mass to $M_a$}
\begin{equation}
m_{\rm max}\equiv \mbox{MIN}(M_a, M_0/2)\,.
\end{equation}
Thus, a simple model for the mass fraction in $(m_a,\delta_a)$-progenitors
to be accreted at $z_a$ may be written as
\begin{eqnarray}\label{eq_calF} 
&& \Phi(s_a,\delta_a \vert S_0,\delta_0; M_a) \nonumber \\
&& = \left\{\begin{array}{l} B^{-1}F(s_a, \delta_a\vert S_0,
    \delta_0; M_a)~~~\mbox{if $m_a\le m_{\rm max}$} \\ 
    0 ~~~\mbox{otherwise}\,,
\end{array}\right. ~~~~~~~~~~~~
\end{eqnarray}
where $F(s_a, \delta_a\vert S_0, \delta_0; M_a)$ is the mass fraction
in progenitor halos of mass $m_a$ to be accreted at redshift $z_a$.
The normalization factor,
\begin{equation}
B=\int_{S(m_{\rm max})}^\infty F(s_a, \delta_a\vert S_0, \delta_0; M_a){\rm d}
s_a \,,
\end{equation}
is the total mass fraction of all progenitors that can be accreted
into the main branch.  Now suppose that the distribution of $M_a$ at
$z_a$ is given by $P(M_a\vert S_0,\delta_0)$ then
\begin{eqnarray}\label{eq:cmfmean}
&& {\cal F}(s_a, \delta_a\vert S_0, \delta_0; {\overline M}_a) \nonumber \\
&& =\int \Phi(s_a,\delta_a\vert S_0,\delta_0; M_a)
P(M_a\vert  S_0,\delta_0) {\rm d} M_a\,.~~~~~~~~~~~~~~~~
\end{eqnarray}

\subsection{Models for $F(s_a,\delta_a\vert S_0,\delta_0; M_a)$}
\label{sec:model_F}

\begin{figure*}
\plotone{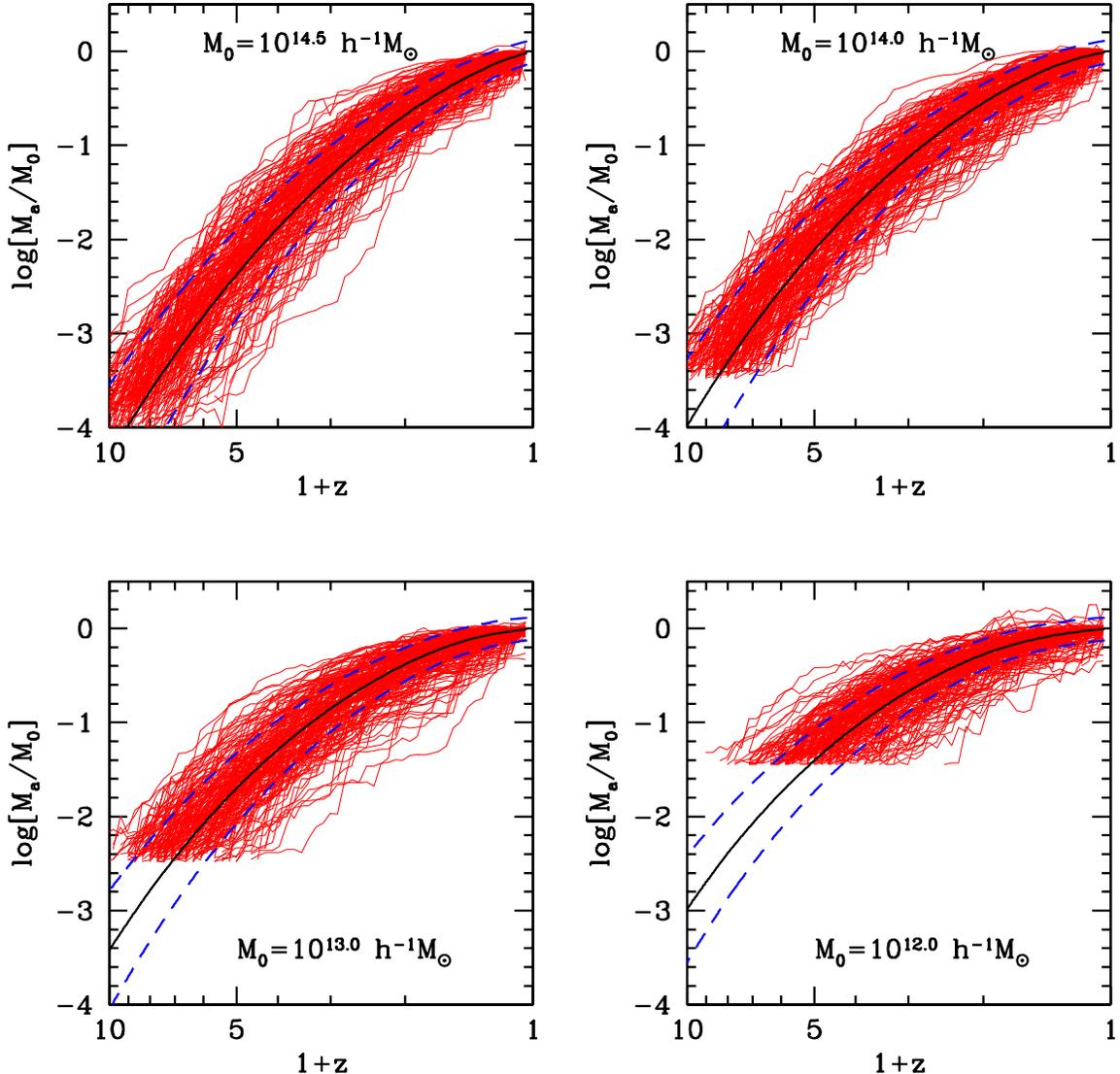}
\caption{The thin solid curves in each panel are the assembly histories of 200
  simulated halos of a given final  mass as indicated in the panel.  The solid
  curve is the predicted median  assembly history by the \citet{Zhao09} model.
  The  two  dashed  curves   are  the  $\pm1\sigma$-range  given  by  equation
  (\ref{eq:sigma1}). }
\label{fig:Assembly}
\end{figure*}

\noindent

{\bf Model I:} According to the extended PS formalism, the fraction of
mass of halo $(M_1, z_1)$ that is in progenitor halos of mass $M_2$ at
redshift $z_2>z_1$ can be written as
\begin{equation}\label{eq:f_1}
f(S_2, \delta_2\vert S_1, \delta_1) 
={1\over \sqrt{2\pi}}
{\delta_2 -\delta_1\over (S_2 - S_1)^{3/2}}
\exp\left[-{(\delta_2-\delta_1)^2\over 2 (S_2-S_1)}\right]\,
\end{equation}
\citep[see][]{LC93}.  Thus, the simplest model is to assume $F(s_a,
\delta_a\vert S_0, \delta_0; M_a)= f(s_a, \delta_a\vert S_0,
\delta_0)$.  However, $f(s_a, \delta_a\vert S_0, \delta_0)$ is the
mean for all $(S_0,\delta_0)$-halos, and so such an assumption
completely ignores the scatter in the assembly history, i.e.  the
dependence on $M_a$.  A better approximation is to assume that $F(s_a,
\delta_a\vert S_0, \delta_0; M_a)$ is independent of $S_0$ and
$\delta_0$, i.e. the accretion properties at $z_a$ is determined
entirely by the mass of the main-branch halo at that redshift
regardless of where the halo will end up at $z=0$.  In this case, we
may write
\begin{eqnarray}
&& F(s_a, \delta_a\vert S_0, \delta_0; M_a) \,{\rm d}\ln \delta_a \nonumber\\
&&~~~~~= f[s_a,\delta_a+{\rm d} \delta_a\vert S (M_a), \delta_a]\nonumber\\
&&~~~~~= {1\over   \sqrt{2\pi}} {{\rm d}\delta_a\over [s_a- S(M_a)]^{3/2}}\,,
~~~~~~~~~~~~~~~~~~~~~~~~~~~~~~~~
\end{eqnarray} 
where the extended  PS formula (\ref{eq:f_1}) is used  in the second equation.
In what  follows we will refer  this model as Model  I.  Note that  as we will
illustrate in section  \ref{sec:CMF}, this model does not  match well with the
simulation results.

\smallskip\noindent 

{\bf Model II:} Numerical simulations have shown that the PS formula
(\ref{eq:f_1}) is not accurate.  Attempts have been made to come up
with better approximations \citep[e.g.][]{ST04, Neistein08, Cole08,
  PCH08}.  According to the empirical modification proposed by
\citet[][]{PCH08}, we may write
\begin{eqnarray}
&& F(s_a, \delta_a\vert S_0, \delta_0; M_a)  ~{\rm d}\ln \delta_a \nonumber\\
&&~~~~~ ={1\over \sqrt{2\pi}} {{\rm d}\delta_a\over [s_a- S(M_a)]^{3/2}}
G\left[{\sigma_a\over \sigma(M_a)}, {\delta_a\over \sigma(M_a)}\right]\,, ~~~~~~~
\end{eqnarray}
where $G(x,y) = G_0 x^{\gamma_1} y^{\gamma_2}$, with $G_0=0.57$,
$\gamma_1=0.38$ and $\gamma_2=-0.01$.  This model will be referred to
as Model II in what follows.

\smallskip\noindent 

{\bf Model III:} Other than the above two models, we find by trial and
error that the following simple modification of the extended PS
formula provides a much more {\bf accurate} model for $F$:
\begin{eqnarray}\label{eq:f_2}
&& F(s_a, \delta_a\vert S_0, \delta_0; M_a) ~{\rm d}\ln \delta_a \nonumber\\
&& ~~~~={1\over \sqrt{2\pi}} {\delta_a-\delta_M \over (s_a - S_M)^{3/2}}
\exp\left[-{(\delta_a-\delta_M)^2\over 2 (s_a-S_M)}\right]\,,~~~~~~~~~  
\end{eqnarray}
where $\delta_M$ corresponds to the redshift at which the main branch
has the mass
\begin{equation}
M_{\rm max}=\mbox{MIN}({\cal M}_a + m_{\rm max}, M_0)\,,
\end{equation} 
with ${\cal M}_a$ the {\it median} main-branch mass (not to be confused
with the {\it mean} branch mass ${\overline M}_a$) and 
\begin{equation}\label{eq:S_M}
S_M\equiv \sigma_M^2=\sigma^2(M_{\rm max})\,.
\end{equation}
This model will be referred to as Model III.

\subsection{Halo Assembly History}\label{sec:assembly}

\begin{figure*}
\plotone{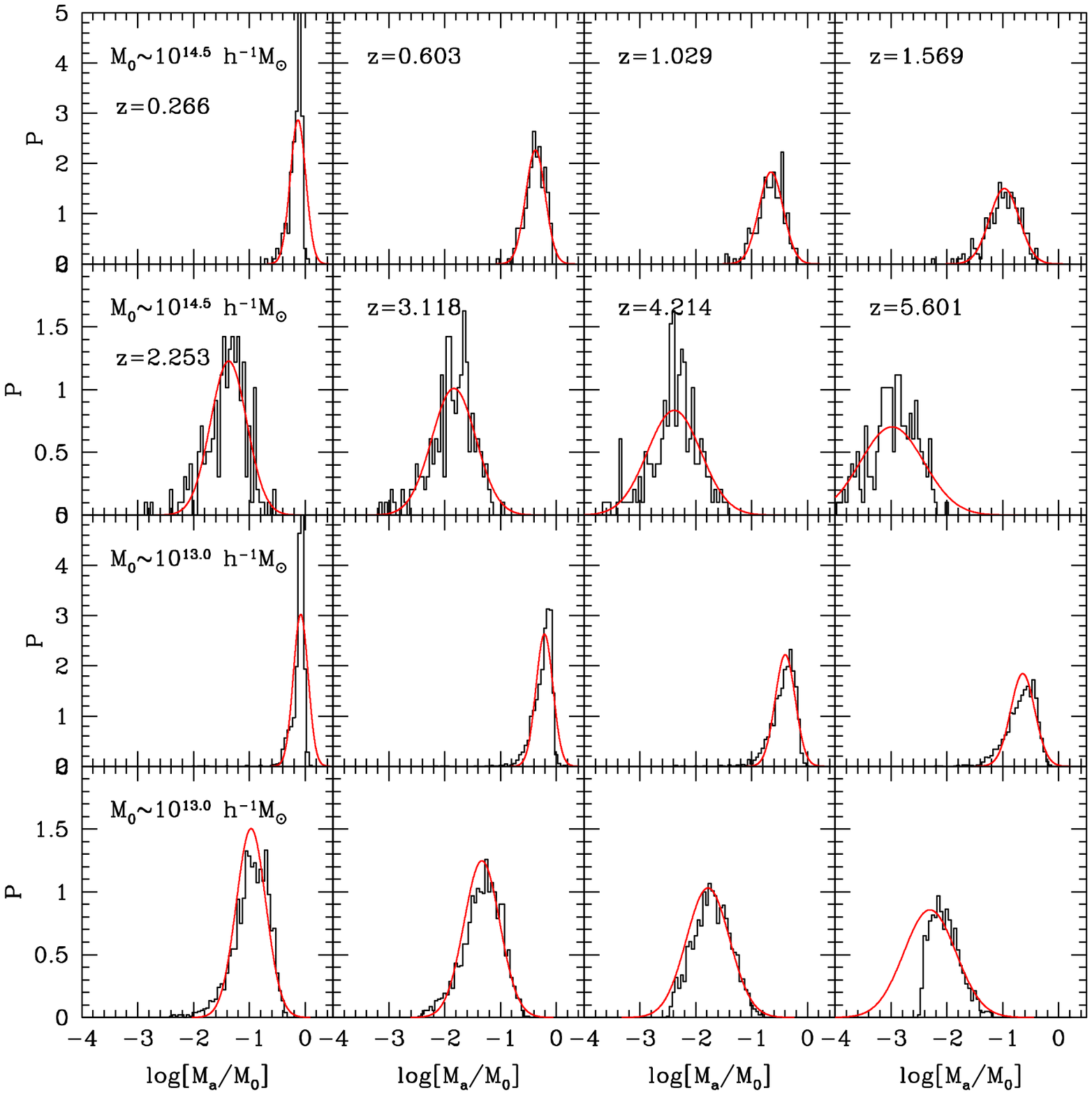}
\caption{The  distribution  of the  main-branch  mass  at different  redshifts
  obtained  from $N$-body  simulations  (histograms).  Results  are shown  for
  halos with  final masses $\sim  10^{14.5}\msunh$ (upper two rows)  and $\sim
  10^{13.0}\msunh$ (lower two rows),  respectively.  The smooth curves are the
  log-normal model described in Section~\ref{sec:assembly}. }
\label{fig:scatter}
\end{figure*}

\begin{figure*}
\plotone{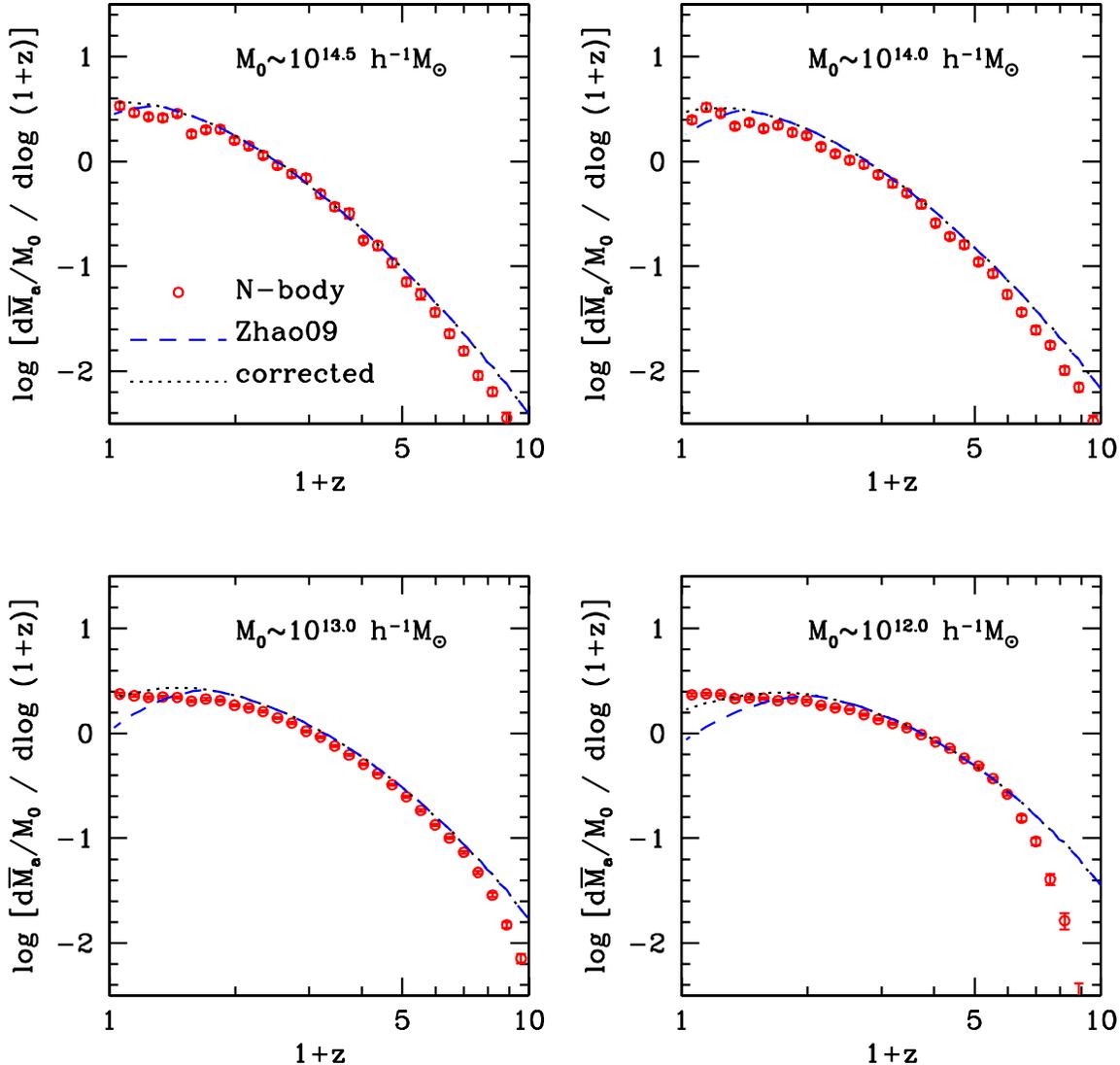}
\caption{The  mean mass  accretion rate  of  host halos  with different  final
  masses.  The  open circles  in each panel  show the  rate in terms  of ${\rm
    d}\log (1+z)$  obtained from $N$-body simulations, normalized  by the mass
  of the host halo, with  error-bars obtained from 200 bootstrap resampling of
  the host halos.  The dashed  line is the model prediction of \citet{Zhao09},
  while the dotted  line shows the corrected mass  accretion rate according to
  equation (\ref{eq:correction}).}
\label{fig:dM_dlogz}
\end{figure*}  

Following \citet{Zhao09}, the median
accretion rate of a $(M_0, \delta_0)$ host halo at redshift $z$ can be
written as
\begin{equation}\label{eq_assembly}
{{\rm d}\ln\sigma (M)\over {\rm d}\ln\delta_c(z)} ={1\over 5.85}
\left\{w[\delta_c(z),\sigma(M)]- p[\delta_c(z);\delta_0,\sigma_0]\right\}\,.
\end{equation}
Here 
\begin{equation}
w={\delta_c(z)\over\sigma(M)} 10^{-{\rm d}\ln\sigma(M)/{\rm d}\ln M}\,;
\end{equation}
and 
\begin{eqnarray}\label{eq:p_zhao}
&& p[\delta_c(z);\delta_0,\sigma_0] \\
&& =p(\delta_0;\delta_0,\sigma_0) \times
  \mbox{Max}\left[0, 1-{\log\delta_c(z)-\log\delta_0\over
      0.272/w(\delta_0,\sigma_0)}\right]\,,~~~ \nonumber
\end{eqnarray}
with 
\begin{equation}
p(\delta_0;\delta_0,\sigma_0) ={1\over 1+[w(\delta_0, \sigma_0)/4]^6}
{w(\delta_0,\sigma_0)\over 2}\,.
\end{equation}
One can obtain the median main-branch mass ${\cal M}_a(z_a|M_0,z_0)$ by simply
integrating Eq.~(\ref{eq_assembly})  from redshift $z_0$ to  $z_a$.  The solid
curves in Fig.~\ref{fig:Assembly} correspond  to the median assembly histories
thus obtained for four different host halo masses, as indicated in each panel.
For comparison, the thin, jagged curves correspond to mass accretion histories
obtained from $N$-body  simulations. The results are shown  for a selection of
200 main branch  assembly histories for halos in each  mass bin extracted from
the $300\mpch$  box simulation.  Clearly,  the model of  \citet{Zhao09} yields
{\it  median} mass  assembly histories  that are  in excellent  agreement with
simulation results.

In order to proceed we need to make one important addition. The \citet{Zhao09}
model only  gives the median assembly  history. However, a  complete model for
the un-evolved subhalo distribution with  respect to the mass at accretion and
the redshift of accretion  requires the full distribution function $P(M_a\vert
S_0,\delta_0)$.  It is  easy to understand that the  dispersion in $M_a$ plays
an important role. Since by definition  the masses of subhalos are all smaller
than that of the main progenitor,  using the mean assembly history would imply
that at  any given time no  accreted subhalo can  have a mass larger  than the
mean mass  on the main  branch. Clearly, when  allowing for dispersion  in the
main-branch  masses, this  constraint  is no  longer  present.  The  different
panels  of Fig.~\ref{fig:scatter}  show the  distributions of  the main-branch
halo  masses at  different redshifts  obtained from  simulations (histograms).
Results are shown  for halos with redshift zero  masses $\sim 10^{14.5}\msunh$
(panels in the upper two rows) and $\sim 10^{13.0}\msunh$ (panels in the lower
two rows), respectively.  All  the distributions are reasonably well described
by a log-normal  distribution, with median given by  the \citet{Zhao09} model,
and with a dispersion (in 10-based logarithm) given by 
\begin{equation}\label{eq:sigma1}
\sigma = 0.12 - 0.15\,\log({\cal M}_a/M_0)\,,
\end{equation} 
as shown by the solid curves  in the figure.  The lack of low-mass main-branch
progenitors  in the  simulation apparent  in the  bottom right  two  panels is
simply due to the mass limit  in the $N$-body simulation.  For comparison, the
dashed lines in Fig.~\ref{fig:Assembly} show  the $\pm 1\sigma$ range given by
Eq.~(\ref{eq:sigma1}), together  with the  median given by  the \citet{Zhao09}
model (solid curves).

An important advantage of the log-normal form for the distribution $P(M_a\vert
S_0,\delta_0)$  is that  it is  straightforward to  compute the  {\it average}
accretion   rate   ${\rm  d}   {\overline   M}_a/{\rm   d}\ln  (1+z_a)$   (see
Eq.~[\ref{eq:N_a_1}]). After  all, for  a log-normal distribution  the average
${\overline  M}_a$  is  related  to  the  median  ${\cal  M}_a$  according  to
${\overline M}_a  = e^{(\ln(10)\sigma)^2/2} \,  {\cal M}_a$. Thus,  assuming a
log-normal distribution,  we can simply  use the \citet{Zhao09} model  for the
{\it  median}  assembly history  to  obtain  the  {\it mean}  accretion  rate.
Fig.~\ref{fig:dM_dlogz}  shows the  mass accretion  rate ${\rm  d} ({\overline
  M}_a/M_0)/{\rm  d}\log (1+z)$ for  halos with  different final  masses.  The
dashed curve  in each panel is  the prediction of the  \citet{Zhao09} model of
the median together with the log-normal distribution described above.
 
In the  simulations, especially  for small halos  at low redshifts,  there are
time-steps over which the accretion rate is negative.  This can come about due
to, for instance, tidal stripping by neighboring structures, the loss of
unbound  subhalos  or  unbound  particles,  the fragmentation of
halos,  and  the FOF-bridging  problem.   
The  \citet{Zhao09}  model  has  taken  such  effects
explicitly into account via a correction factor in equation (\ref{eq:p_zhao}).
In general, however,  it is possible to have two  different definitions of the
un-evolved subhalo population, one based on all subhalos that have entered the
main branch at some point (but  are not necessarily bound to or located within
the final halo), and  the other based on those that have  more than
half of their particles  ending up in  the final  halo.  In  this paper  we adopt  the first
definition in which the accretion rate is determined by all subhalos that have
been  accreted onto  the main  branch  at some  point in  time, regardless  of
whether  they have  left the  main  branch again  or not.   The open circles  in
Fig.~\ref{fig:dM_dlogz} show the simulation  results based on this definition.
As one can see, the mean  mass accretion rates predicted by the \citet{Zhao09}
model match  well the simulation results  over a large range  of redshift; the
mismatch seen at high-$z$ for low-mass  halos is simply an artefact due to the
limited  numerical  resolution  of   the  simulation.   The  only  significant
discrepancy occurs for low-mass halos at low redshift where the predicted rate
is lower than the simulated rate.   This discrepancy is mainly due to the fact
that  the \citet{Zhao09}  model  tries to  account  for the  affects of  tidal
stripping and  the presence  of unbound subhalos.   For our definition  of the
un-evolved subhalo  population, however, such correction is  not required when
computing  ${\rm  d}  {\overline  M}_a/{\rm  d}\ln (1+z)$.   As  discussed  in
\citet{Zhao09}, this can be achieved by simply setting %
\begin{equation}\label{eq:correction}
p[\delta_c(z);\delta_0,\sigma_0]=0
\end{equation}
in Eq.~(\ref{eq_assembly}).  The dotted lines in Fig.~\ref{fig:dM_dlogz}
show the predictions based on Eq.~(\ref{eq:correction}).  Clearly this
simple modification works remarkably well, bringing the predictions in
good agreement with the simulation results.

If one  is interested  in using  the second definition  for the  population of
subhalos,  one can  still  use our  model  but with  Eq.~(\ref{eq:correction})
replaced by Eq.~(\ref{eq:p_zhao}). In  what follows, we present predictions of
Model    III   based    on    the   first    definition,   corresponding    to
Eq.~(\ref{eq:correction}).  We  have tested,  though,  that  using the  second
definition  instead yields  results  that  are very  similar,  except when  it
concerns the subhalos accreted at low redshifts in low-mass host halos.

\section{Test with $N$-body Simulations}
\label{sec_test}

In  this section  we  use the $N$-body  simulations described 
in Section 2  to test  our  models for  the
distribution of  subhalos with respect  to their accretion redshift  and their
mass  at accretion.  
   
\begin{figure*}
\plotone{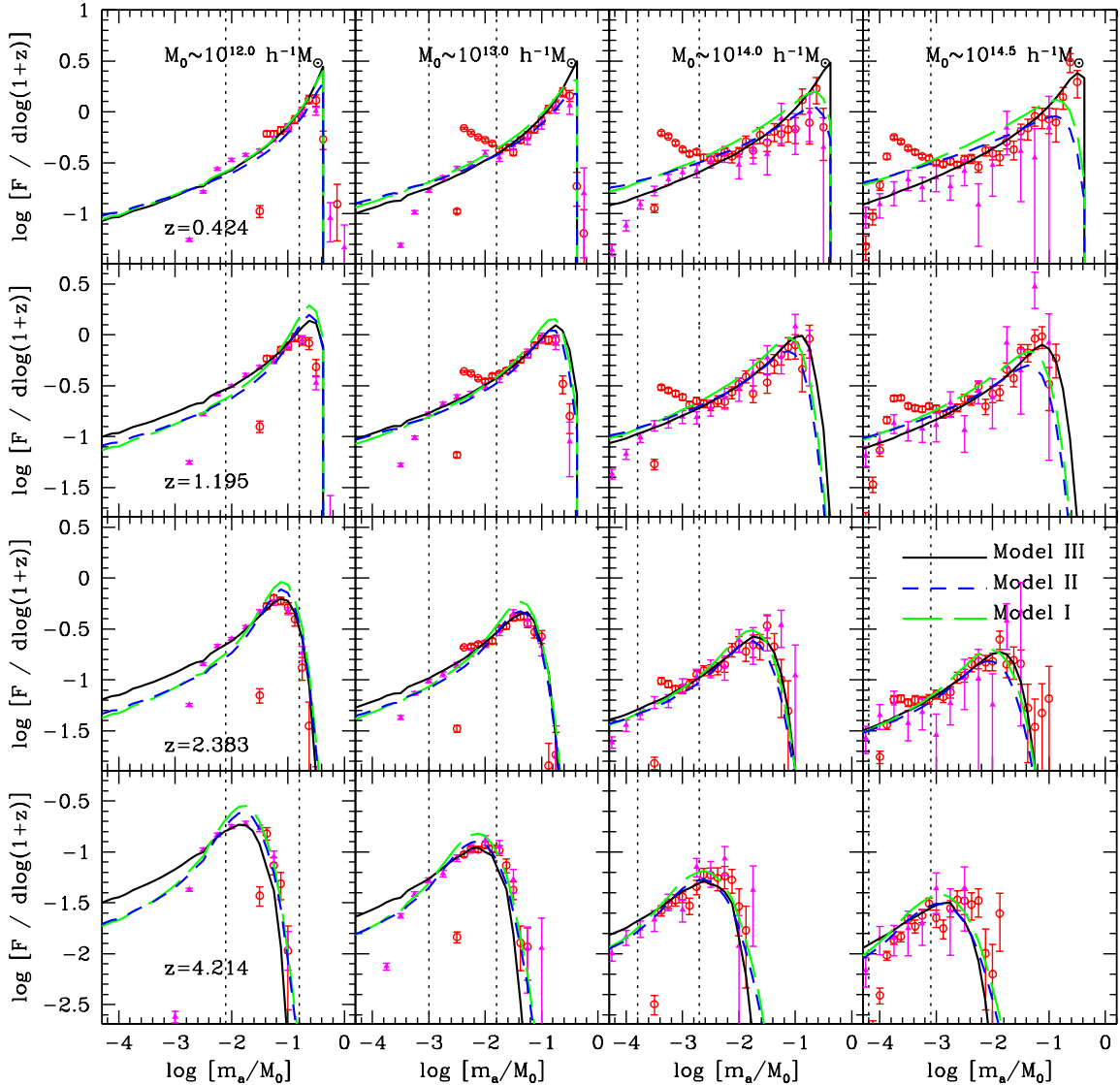}
\caption{The  conditional  mass distribution  of  subhalos  in  host halos  of
  different  masses  (different  columns)  at different  redshifts  (different
  rows).   The symbols are  results from  N-body simulations,  with error-bars
  obtained from  200 bootstrap resampling of  the host halos.   In each panel,
  results are shown for both high- ($100\mpch$ box; filled triangles) and low-
  ($300\mpch$   box;   open   circles)   resolution   simulations.    In   the
  high-resolution  simulation, unbound  particles  are removed  (see text  for
  details).  The long-dashed,  dashed and solid lines show  the predictions of
  Models I, II,  and III respectively.  Finally, the  vertical dotted lines in
  each panel correspond to the mass  limit of 100 particles in the high- (left
  line) and low- (right line) resolution simulations, respectively.  Note that
  bin widths in different columns are different. }
\label{fig:mass_dist}
\end{figure*}

\begin{figure*}
\plotone{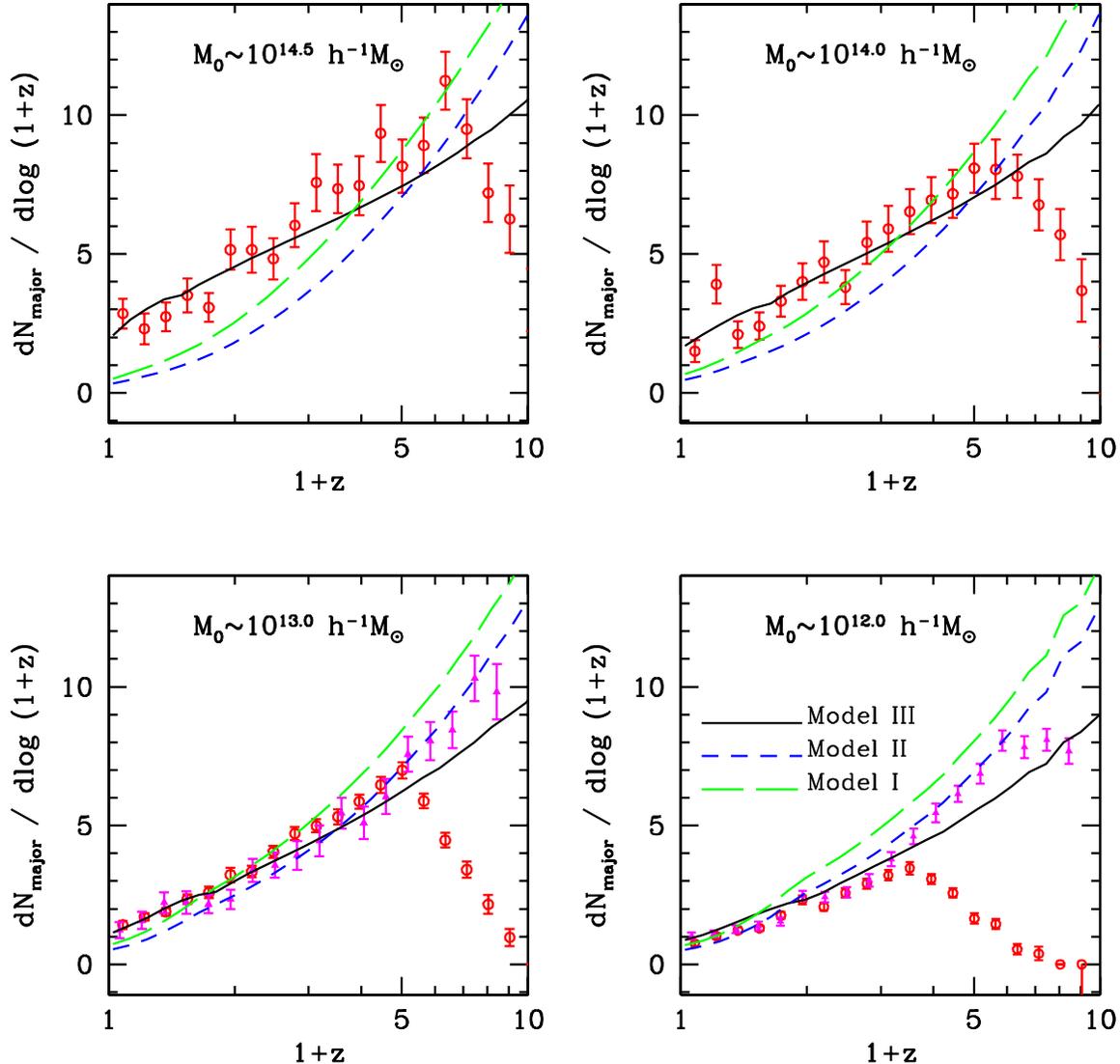}
\caption{The  major  merger events  per  unit  $\log(1+z)$  as a  function  of
  $z$. The  symbols show  results obtained from  simulations, while  the three
  curves in  each panel  show the predictions  of Models  I, II and  III.  For
  comparison, results  for both the  high-resolution (which extends  to higher
  redshift;  filled triangles) and  low-resolution (open  circles) simulations
  are shown for cases where the the statistic is sufficiently good.}
\label{fig:major}
\end{figure*}

\begin{figure*}
\plotone{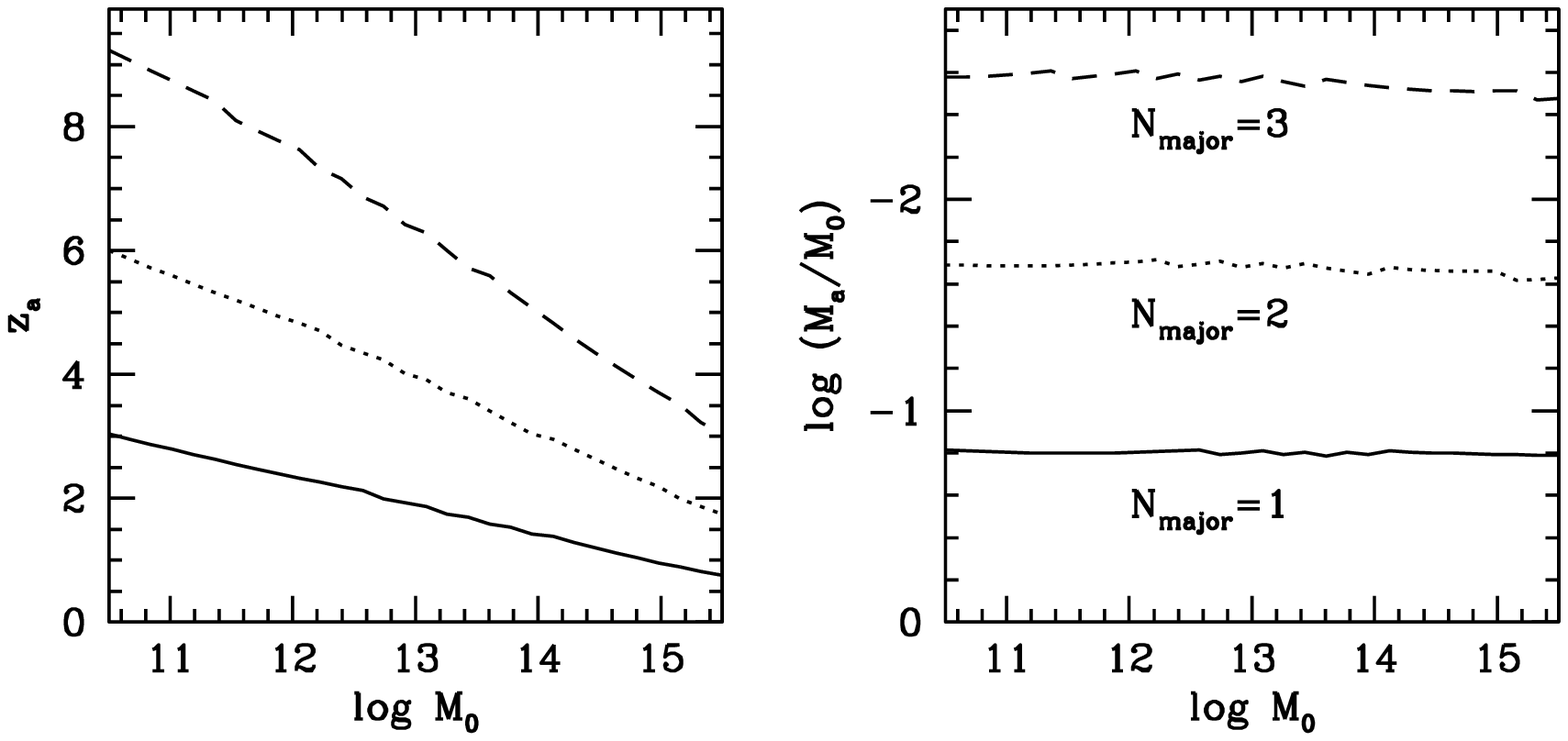}
\caption{The average redshifts  below which (left panel) and  the average main
  progenitor to host halo mass  ratios above which (right panel) that subhalos
  have experienced  1st (solid), 2nd  (dotted), 3rd last (dashed  lines) major
  merger events, predicted by Model III. }
\label{fig:majorN}
\end{figure*}

\subsection {The conditional mass function}\label{sec:CMF}
 
The first quantity we consider is the conditional mass function of subhalos in
host halos of different masses (Eq. [\ref{eq:cmfmean}]).  From the halo merger
trees constructed  from the simulations,  we measure the mass  distribution of
accreted subhalos  within equal $\log  (1+z)$-bins centered at  $0.42$, $1.2$,
$2.4$   and   $4.2$,   respectively.    The   results  are   shown   in   Fig.
\ref{fig:mass_dist} as open circles (for $300\mpch$ simulation box) and filled
triangles  (for  $100\mpch$ simulation  box)  where  the  errorbars have  been
obtained from  200 bootstrap resamples of  the population of  host halos. Each
column shows  results for host halos with  a given mass, while  each row shows
the results  at a given redshift.   Note that, because of  mass resolution and
because  of tidal effects,  a fraction  of the  subhalos containing  $\la 100$
particles, especially those near massive halos, are not gravitationally bound.
This is the reason for the  artificial upturn in the conditional mass function
seen  in  the  low-resolution  simulation  results at  low  redshift.  In  the
literature  this problem  has been  treated either  by using  only  halos that
contain  large  enough  number of  particles  or  by  getting rid  of  unbound
particles  from halos.   For instance,  the SUBFIND  halo finder  developed by
\citet[][]{Sp01b} tries to separate bound and unbound structures in FOF halos.
In \citet[][]{Benson01},  unbound particles are removed  iteratively from each
halo until the total energy of the halo becomes negative. In order to quantify
the magnitude of this effect, we have also constructed halo merger trees using
a  method  similar  to  \citet{Benson01}  to calculate  the  number  of  bound
particles  per  (sub)-halo.  Halos  with  less than  20  bound  particles  are
discarded from our  sample. In Fig.  \ref{fig:mass_dist}, we  show the results
obtained from the $100\mpch$ simulation box in which we removed the halos with
bound particles less than 20 (using the method described above), together with
the results  from the $300\mpch$  simulation box without such  treatment.  The
dotted, vertical lines  correspond to halos with 100 particles  in each of the
two simulations, and  are show for comparison.  As one  can see, the treatment
of the  unbound halos is  able to remove  the spurious upturn at  the low-mass
end. But for sure and to  be conservative, the results presented below are all
for subhalos containing at least 100 particles.

The  long dashed  lines in  Fig.~\ref{fig:mass_dist} show  the  predictions of
Model I.   This model does  not match very  well with the  simulation results,
especially for  massive halos at low  redshifts.  As shown  in the upper-right
panel  of   Fig.~\ref{fig:mass_dist},  for  massive  host   halos  this  model
significantly  over-predicts  the number  of  accreted  low-mass subhalos  and
under-predicts  that  of  accreted  massive subhalos.   Such  discrepancy  has
already    been   noticed    and   extensively    discussed    in   literature
\citep[e.g.][]{ST02, Cole08, PCH08, Neistein10}. An empirical modification was
suggested  by \citet{PCH08}, which  is adopted  in our  Model II.  The results
obtained from this  model are shown in Fig.~\ref{fig:mass_dist}  as the dashed
lines.   As one can  see, this  modification successfully  suppresses low-mass
subhalos, but makes the  under-prediction of massive subhalos worse.  Finally,
let  us  look   at  our  Model  III,  the  results  of   which  are  shown  in
Fig.~\ref{fig:mass_dist}  as solid  curves.   Clearly, Model  III matches  the
simulation  results much  better than  either model  I or  II,  especially for
massive hosts.

Note   that  all   three  models   shown  have   adopted  the   correction  of
Eq.~(\ref{eq:correction}). If  we use model  III without this  correction, the
model  underpredicts the  abundance of  subhalos (as  defined using  the first
definition  described in \S\ref{sec:assembly}),  especially for  low-mass host
halos  at low  redshifts.  For completeness,  we  have also  tested model  III
without the  dispersion in  halo assembly histories.  As expected,  not taking
this dispersion into account yields conditional subhalo mass functions in poor
agreement  with the  simulation,  especially at  higher  redshifts. Hence,  we
caution that  it is important to  properly account for the  dispersion in halo
assembly histories.

\subsection {Redshift distribution of major mergers}

With the conditional mass functions  described above, it is straightforward to
calculate the accretion  rate of subhalos as a function of  host halo mass and
redshift. Here we  focus on the rates  of major mergers where the  mass of the
accreted subhalo is required to be  $m_a\ge M_a/3$. This mean rate in terms of
redshift interval can be written as 
\begin{equation}\label{eq:major}
{{\rm d} N_{\rm major}\over {\rm d}\ln (1+z_a)} 
= \int_{M_a/3}^{M_a} {\cal N}_a (s_a, \delta_a\vert S_0, \delta_0) {{\rm d} m_a
  \over m_a} \,.
\end{equation}
Note that since the major mergers  are defined with respect to individual main
branch  masses   $M_a$,  here  we   do  not  include  the   lognormal  scatter
(Eq. (\ref{eq:sigma1})  ) in performing  the integration.  The  predictions of
Models I,  II and III are shown  in Fig.  \ref{fig:major} as  the long dashed,
dashed  and solid  lines, respectively.   To  check the  model predictions  we
calculate the  same quantity  directly from the  $N$-body simulations  and the
results are  shown in Fig.  \ref{fig:major}  as symbols with  error bars.  For
comparison,  we  show  the  results  obtained from  both  the  $300\mpch$  and
$100\mpch$ simulation  boxes.  In  the case of the  more massive halos  (upper two
panels), however,  we do not show  the results from  the $100\mpch$ simulation
box,  since  these  are  very  noisy  due to  small  number  statistics.   The
$300\mpch$  box,  although having  better  statistics,  has insufficient  mass
resolution  to properly resolve  the merger  statistics of  low mass  halos at
higher redshifts (as is clearly  evident from the lower two panels).  Clearly,
the  results of  model III  are  in very  good agreement  with the  simulation
results,  and fair  much better  than either  Model I  or Model  II.   In what
follows, we focus only on the predictions of Model III.

From the model for the major merger  rates in halos of different masses and at
different redshifts, we  can also predict the characteristic  redshift and the
characteristic  mass associated with  the last  major merger.   By integrating
Eq.~(\ref{eq:major}) over redshift  from $0$ to some $z_a$,  we can obtain the
average number  of major mergers,  $N_{\rm major}$, expected in  this redshift
interval.  We  call the  value of $z_a$  the characteristic redshift,  and the
main-branch mass at this redshift  the characteristic mass of the last $N_{\rm
  major}$th   major   mergers.    In   Fig.~\ref{fig:majorN}   we   show   the
characteristic redshift (left panel)  and characteristic mass (right panel) as
functions of host-halo mass, for  $N_{\rm major}=1$, 2 and 3, corresponding to
the  1st, 2nd and  3rd last  major merger,  respectively.  Since  more massive
halos assemble  later (see Fig.\ref{fig:Assembly}), their  last major mergers,
on average, occur at lower redshifts,  as shown clearly in the left-hand panel
of       Fig.~\ref{fig:majorN}       (see       similar      findings       in
\citet[][]{Li07}). Interestingly, the characteristic  mass of those last major
mergers, in units of the host  halo mass, is virtually independent of the host
halo mass.  This is a manifestation  of the fact  that all halos (in  the mass
range explored  here) have average  merger histories that are  self-similar if
they  are allowed  to  be stretched  or  shortened along  the time-axis.   For
example, it  is interesting to notice  that the average  time interval between
major mergers  is roughly the  same as the  time during which the  main branch
mass increases  by a constant  factor, about $10^{0.8}$, quite  independent of
the host halo mass.

In a recent paper based on the Millennium simulations, \citet{FMB10} have come
up with a fitting formula which describes the subhalo accretion rates obtained
from the  simulations. However,  since their fitting  formula is not  based on
cosmology-independent  quantities,  it  is   only  valid  for  the  particular
cosmology    adopted   in    the   Millennium    simulations    (see   Section
\ref{sec_universal} for more details).

\subsection {Redshift distribution of subhalo accretion}         

\begin{figure*}
\plotone{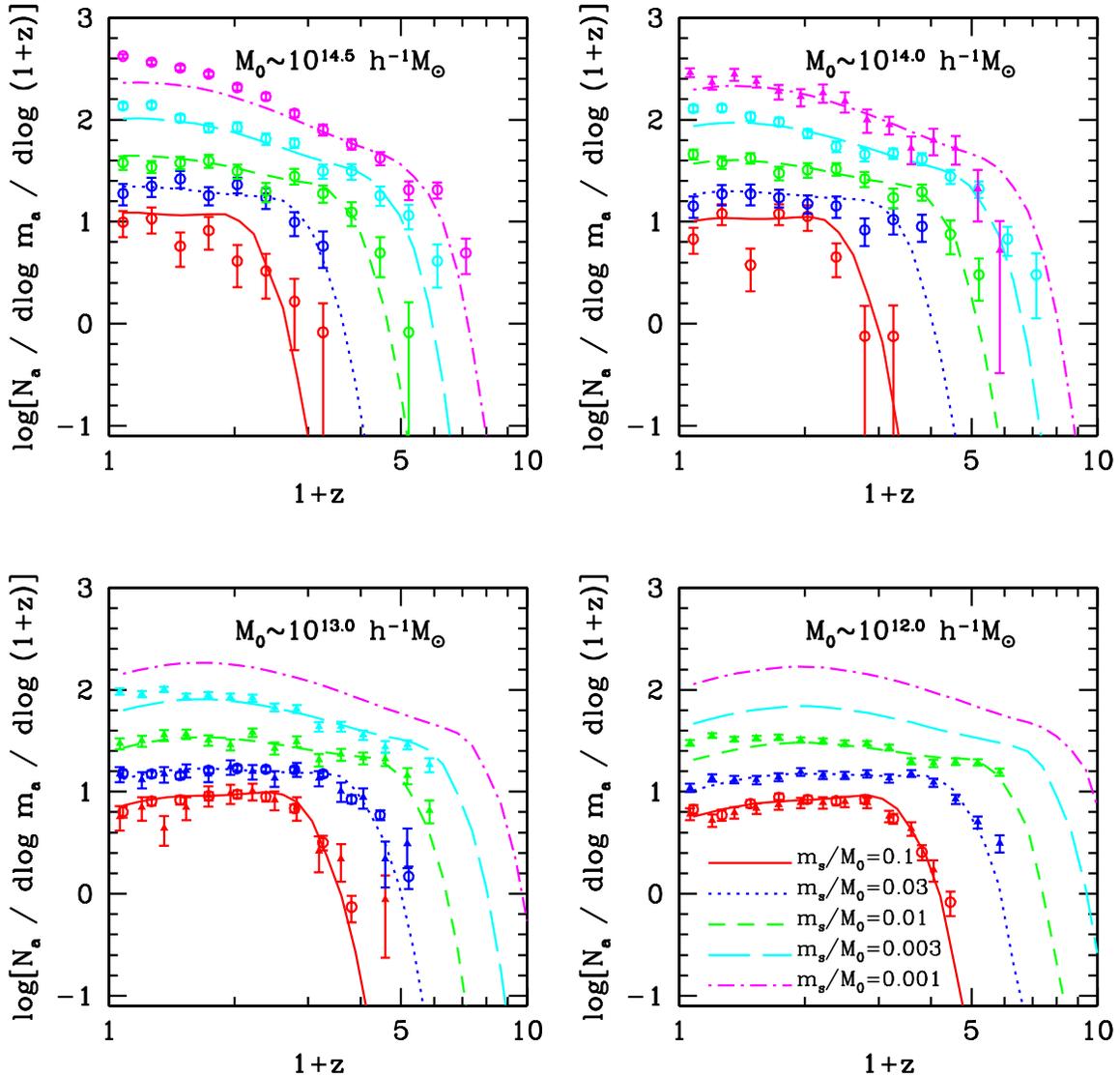}
\caption{Model  predictions for  the distribution  of accretion  redshifts for
  subhalos  with $m_a/M_0=0.1$  (solid lines),  $0.03$ (dotted  lines), $0.01$
  (dashed lines),  $0.003$ (long dashed lines) and  $0.001$ (dot-dashed lines)
  respectively.   Results are  shown for  host  halos of  different masses  as
  indicated in the  panels.  These results assume a  $\Lambda$CDM universe and
  are  compared with  the results  obtained from  the $300\mpch$  box $N$-body
  simulations with the same cosmology (open circles).  For comparison, results
  obtained  from the  $100\mpch$ box  simulations  are also  shown (as  filled
  triangles) for cases where statistics are sufficiently good.}
\label{fig:7}
\end{figure*}
\begin{figure*}
\plotone{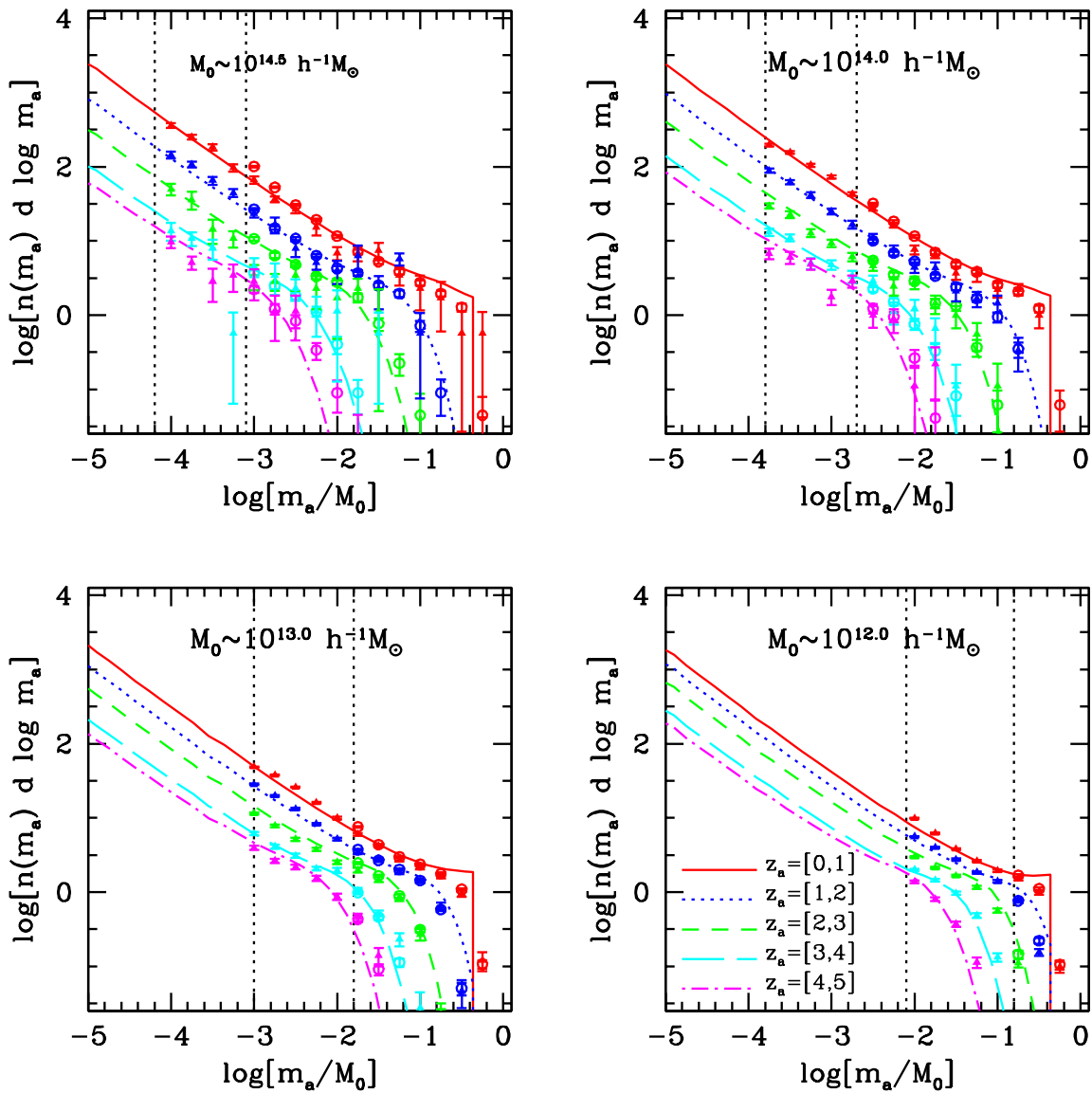}
\caption{The un-evolved mass function of subhalos accreted within the redshift
  ranges [0, 1] (solid lines), [1,  2] (dotted), [2, 3] (dashed), [3, 4] (long
  dashed)  and [4, 5]  (dot-dashed), for  host halos  of different  masses, as
  indicated  in each panel.   Here model  predictions assuming  a $\Lambda$CDM
  cosmology are compared with the the results obtained from the $300\mpch$ box
  $N$-body simulations of the  same cosmology (open circles).  For comparison,
  results  obtained from  the $100\mpch$  box  simulation are  also shown  (as
  filled triangles) for cases where the statistics are sufficiently good.  The
  vertical lines in each panel correspond  to the mass of 100 particles in the
  two simulations, as in Fig. \ref{fig:mass_dist}).}
\label{fig:8}
\end{figure*}
\begin{figure}
\plotone{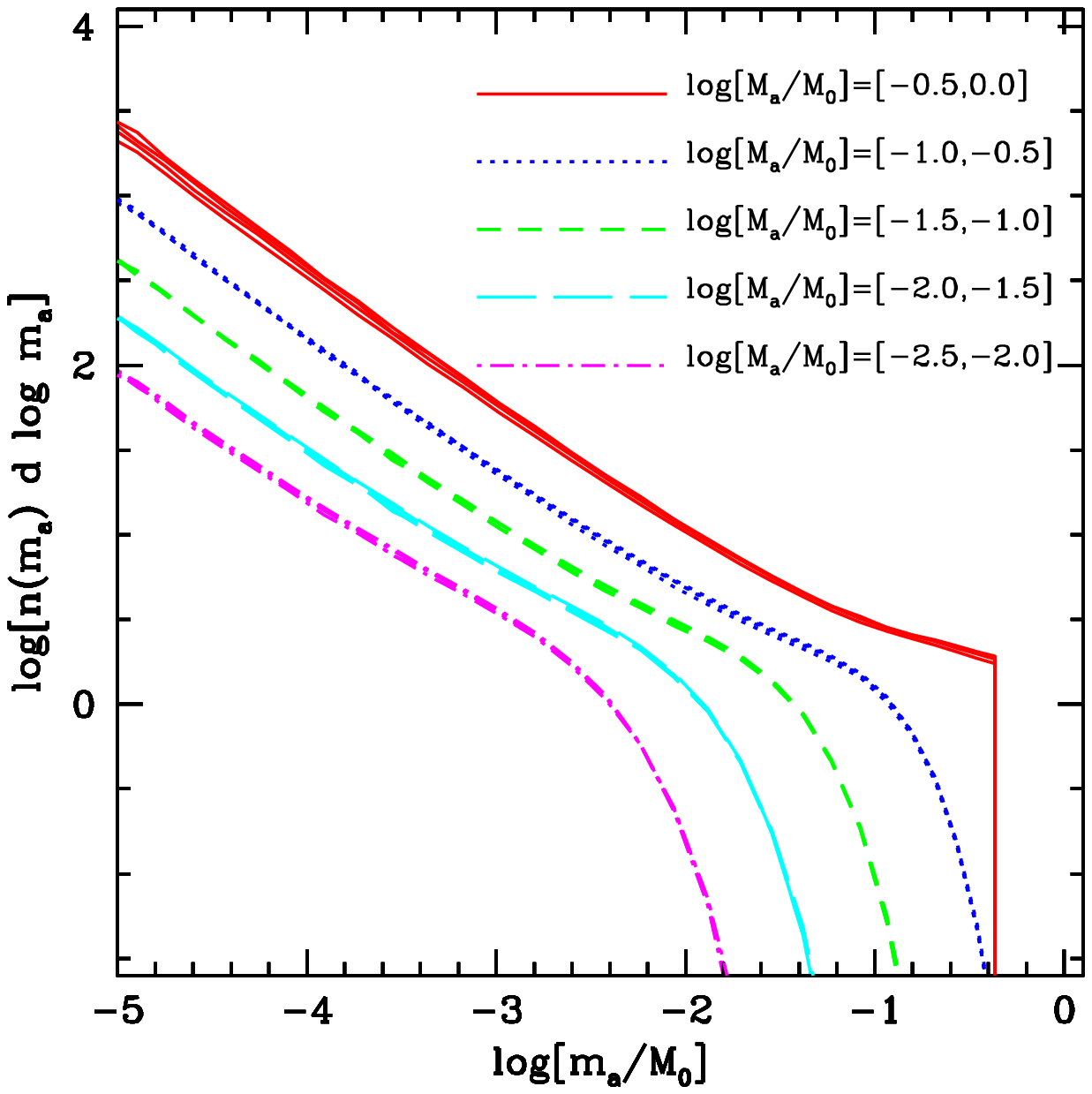}
\caption{The un-evolved  mass function  of subhalos accreted  within different
  $\log  [{\cal M}_a/M_0]$  
ranges  [-0.5, 0.0]  (solid  lines), [-1.0,  -0.5]
 (dotted), [-1.5, -1.0] (dashed), [-2.0, -1.5] (long dashed) and [-2.5, -2.0]
 (dot-dashed). For each range of  $\log  [{\cal M}_a/M_0]$,   
results  are shown  with the same line style for  host halos  of four  
different masses: $M_0\sim 10^{14.5}$, $10^{14.0}$,
$10^{13.0}$, and $10^{12.0}h^{-1}{\rm M}_\odot$. Note that shown 
in this way the results for different $M_0$ are almost indistinguishable. }
\label{fig:9}
\end{figure}

In Fig. \ref{fig:7} we show our  model predictions for the distribution of the
accretion  redshift  for subhalos  with  $m_a/M_0=0.1$  (solid lines),  $0.03$
(dotted  lines) ,  $0.01$  (dashed  lines), $0.003$  (long  dashed lines)  and
$0.001$  (dot-dashed lines),  respectively.   The results  are obtained  using
Eq. \ref{eq:N_a_1}  by making some simple  coordinate transformations.  Results
shown  in  different  panels  are  for  host halos  of  different  masses,  as
indicated.  Open  circles and filled  triangles indicate the  results obtained
from the $300\mpch$ and  $100\mpch$ simulations boxes, respectively, where the
error-bars  have been  obtained using  200 bootstrap  resamples.   The various
lines  show  the  predictions  based  on  Model III,  and  overall  match  the
simulation  results remarkably  well.  Note  that the  accretion  rate depends
strongly on the mass  of the host halo.  For the same  mass ratio, subhalos in
more massive hosts  are accreted later, reflecting the  hierarchical nature of
structure formation in the $\Lambda$CDM cosmology.

\subsection {Un-evolved subhalo mass functions}

Finally, let us look at the un-evolved subhalo mass functions.  By integrating
Eq.~(\ref{eq:d2N_a1})  over  a  given   redshift  range,  we  can  obtain  the
un-evolved mass function of the  subhalos accreted in that redshift range.  In
Fig. \ref{fig:8} we show the un-evolved mass functions of subhalos accreted in
the redshift ranges [0,  1], [1, 2], [2, 3], [3, 4]  and [4, 5], respectively.
Results are  shown for host  halos of different  masses, as indicated  in each
panel.  Here  again, symbols indicate  the results from our  simulation boxes,
while  lines show  the predictions  of Model  III.  Clearly,  our model  is in
excellent agreement with  the simulation results at all  redshifts and for all
host masses. Upon close inspection,  it is clear that the un-evolved subhalo
mass function for a given redshift  range depends on host halo mass, especially
at high redshift: in terms of  the scaled mass, $m_a/M_0$, the subhalo mass
function  at high  $z$  is  significantly higher  for  lower-mass host  halos.
Moreover,  the  normalization   of  the  un-evolved  subhalo  mass
function at a given redshift for halos of different masses seem to
be roughly proportional to the assembly  history of the host halos 
shown in Fig.   \ref{fig:Assembly}.  To  test this,  we show  in Fig.   \ref{fig:9} the
un-evolved subhalo mass  functions for different  host  halos
at the time when the host halos have assembled a fixed fraction 
of their final masses, i.e. for subhalos accreted  in a given range 
of $\log  [{\cal M}_a/M_0]$  range. Results are shown for five 
different ranges of $\log  [{\cal M}_a/M_0]$: $[-0.5, 0.0]$  (solid
lines), $[-1.0,  -0.5]$ (dotted), $[-1.5, -1.0]$ (dashed), 
$[-2.0, -1.5]$ (long dashed), and $[-2.5, -2.0]$ (dot-dashed).   
For each range of  $\log  [{\cal M}_a/M_0]$  the subhalo mass 
functions for four host halo masses, $M_0\sim 10^{14.5}$, $10^{14.0}$,
$10^{13.0}$, and $10^{12.0}h^{-1}{\rm M}_\odot$, are plotted with the 
same line style. Interestingly, the mass function of subhalos 
accreted in a given range of $\log [{\cal M}_a/M_0]$ is almost 
independent  of the host halo mass, demonstrating that the 
amplitude of of the un-evolved subhalo mass function at a 
given redshift is directly related to the mass assembly rate
of the host halo at that redshift  \citep[see also][]{Giocoli10}.

\begin{figure*}
\plotone{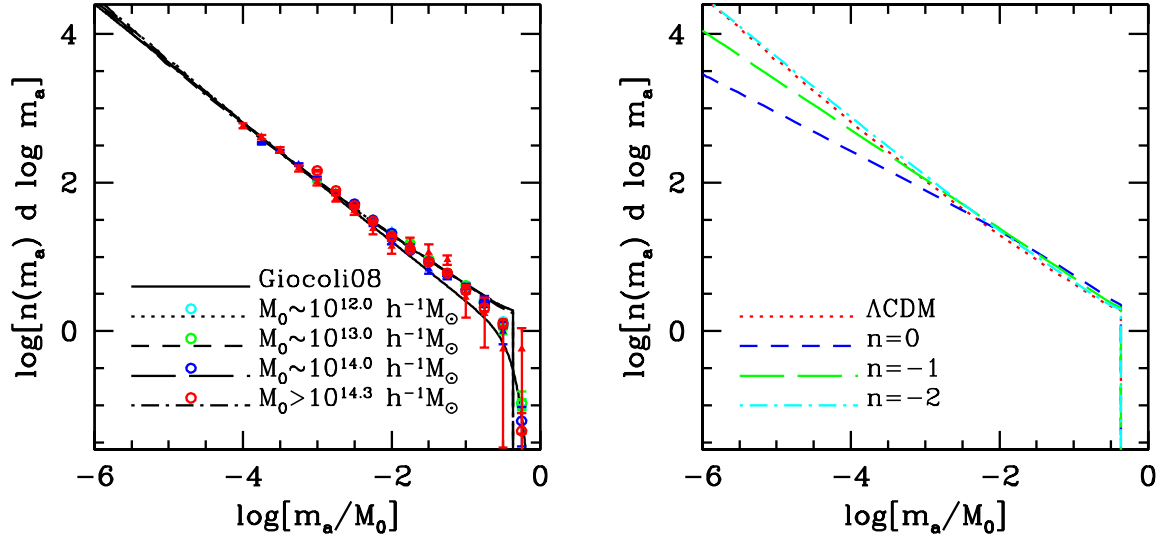}
\caption{Left panel: the un-evolved subhalo mass distribution as a function of
  $m_a/M_0$ for  host halos  of different final  masses at $z=0$.   Here model
  predictions (lines)  are compared  with the results  obtained from  both the
  $300\mpch$ box (open circles) and $100\mpch$ box (filled triangles) $N$-body
  simulations of  the same cosmology. For  comparison, we also  include in the
  figure the  fitting formula obtained  by \citet{Gioc08a} as the  solid line.
  Right panel: The predicted  un-evolved subhalo mass functions for scale-free
  models with  $n=0$, $-1$ and $-2$.   For comparison, we also  include in the
  figure the  model prediction for  the $\Lambda$CDM model considered  in this
  paper.  }
\label{fig:10}
\end{figure*}

Integrating Eq.~(\ref{eq:d2N_a1}) over the {\it full} redshift range (here for
practice we adopt  redshift range $z=0-10$) yields the  {\it total} un-evolved
subhalo  mass function.  The  results are  plotted in  the left-hand  panel of
Fig.~\ref{fig:10} for host halos of different masses.  Interestingly, although
host halos of different masses  accrete their subhalos at different redshifts,
their total un-evolved  subhalo mass functions are extremely  similar, both in
the simulations (symbols)  and in the model (lines).   Note, though, that this
universality  of the  un-evolved subhalo  mass  function, first  hinted at  in
\citet{vdB05},  is  only  approximate.  This  is  due  to  the fact  that  the
un-evolved  subhalo  mass   function  depends  on  the  {\it   shape}  of  the
perturbation  power spectrum  around  the  mass scale  in  question.  For  the
$\Lambda$CDM cosmology considered  here, the slope of the  power spectrum only
changes by  a small amount  over the mass  range $10^{12}\msunh \lta  M_0 \lta
10^{15}\msunh$.   Indeed,   close  inspection   of  the  left-hand   panel  of
Fig.~\ref{fig:10} reveals  small differences between the  curves for different
host masses.

To further illustrate  the dependence of the un-evolved  subhalo mass function
on the shape of the  power spectrum, the right-hand panel of Fig.~\ref{fig:10}
shows the  model predictions for  the total un-evolved subhalo  mass functions
for scale-free models  with power spectra with spectral  indices $n=0$ (dashed
line), $n=-1$  (long dashed line) and $n=-2$  (dot-dashed line), respectively,
together  with  that  for  $10^{12}h^{-1}{\rm  M}_\odot$  host  halos  in  our
$\Lambda$CDM cosmology (dotted line).  The  prediction for the $n=-2$ model is
very close  to that of  the $\Lambda$CDM model,  reflecting the fact  that the
effective spectral index  of the $\Lambda$CDM power spectrum  is close to $-2$
over the mass scales in question. The subhalo mass functions for the $n=0$ and
$n=-1$ models are much shallower.

Finally,  for   comparison,  the  solid   line  in  the  left-hand   panel  of
Fig.~\ref{fig:10} shows  the fitting formula obtained  by \citet{Gioc08a} from
$N$-body simulations  of the $\Lambda$CDM model.  This  fitting formula agrees
well with our model prediction for  the $\Lambda$CDM model, except at the high
mass end where it is slightly  lower than our model prediction. Compared to the
N-body  simulation results, our model prediction agrees with the data
slightly better, except perhaps at the  very high  massive end.  

\section{Discussion}
\label{sec_discussion}

In this paper  we have developed an analytical model for  the mass function of
CDM subhalos  at their  time of  accretion and for  the distribution  of their
accretion times. This model can be used to predict the un-evolved subhalo mass
function,  the  mass  function of  subhalos  accreted  at  a given  time,  the
accretion-time  distribution of  subhalos of  a  given initial  mass, and  the
frequency of  major mergers as  a function of  time. We have tested  our model
against results obtained from  high-resolution $N$-body simulations, and found
that the  model predictions match  the simulation results extremely  well.  In
this section,  we first  discuss the  universality of our  model based  on the
ingredients used in  the construction of the model.   We then briefly describe
two possible applications of our model.

\subsection{The Universality of the Model}
\label{sec_universal}

The first ingredient of our model is the adopted form for $F(s_a,\delta_a\vert
S_0,\delta_0; M_a)$.   As described  in Section~\ref{sec:model_F}, all  of our
three models, I,  II and III, are  based on the extended PS  theory, which has
been  shown   to  work  reasonably   well  for  various   hierarchical  models
\citep[e.g.][]{LC94,Cole08}.  The modifications made in Models II and III have
not been tested  for other cosmological models.  However,  the fact that these
models  work well  for the  $\Lambda$CDM model  over a  large  redshift range,
during  which the  cosmological  parameters change  by  fairly large  amounts,
suggests that the modifications should also work for other cosmological models
with  similar  power  spectra.   Unfortunately, the  effective  power  indices
covered by the $\Lambda$CDM spectrum  are limited, and it is presently unclear
whether  the modifications  will  work  equally well  for  models with  vastly
different power  spectra. It will be  interesting to test the  validity of our
model using scale-free models that cover a large range in spectral indices.

The second ingredient  of our model is the  form of $P(M_a\vert S_0,\delta_0)$
described   in   Section  \ref{sec:assembly}.    As   tested  extensively   in
\citet{Zhao09}, their model for the  median accretion history is universal and
works not  only for realistic CDM  models but also for  scale-free models with
different spectral indices.   Since our tests cover a  range of redshifts over
which  the cosmological parameters  change by  a large  amount, we  expect the
log-normal form of $P(M_a\vert S_0,\delta_0)$,  and the mass dependence of its
dispersion (Eq.~[\ref{eq:sigma1}]), to also apply  to other models that have a
power-spectrum  that  is not  too  different  from  that of  the  $\Lambda$CDM
cosmology  considered here. Here  again, it  will be  interesting to  test the
validity of our model for other power spectra using scale-free models.
 
Similar    arguments    also    apply    to   the    correction    given    by
Eq.~(\ref{eq:correction}).   This  `correction'  is  empirically  obtained  by
\citet{Zhao09}  from  various CDM  and  scale-free  models.   As discussed  in
\S\ref{sec:assembly},  using  the  model  with or  without  this  `correction'
corresponds to two different definitions of the un-evolved subhalo population,
and only has a non-negligible impact when it comes to subhalos accreted at low
redshifts in low-mass host halos.

To  summarize,  we  believe that  our  model  holds  for  any variant  of  the
$\Lambda$CDM  model.  In  particular,   it  should  work  accurately  for  any
cosmological   model  whose   parameters   are  in   agreement  with   current
observational constraints from the cosmic microwave background, supernovae Ia,
and galaxy redshift surveys.

\subsection{Applications}
\label{sec_application}

Our model has a number of applications.  Here we focus on two of them: (i) the
evolution  of  subhalos  as  they  orbit  within their  hosts,  and  (ii)  the
construction of a self-consistent model to link galaxies and dark matter halos
across cosmic time.

The  model  described  here  applies  to  subhalos  at  accretion,  i.e.   the
un-evolved population  of subhalos. However, as  a small halo  merges with and
orbits  in a  larger halo,  it is  subjected to  tidal forces  from  the host,
causing it  to lose mass, and to  dynamical friction, which causes  it to lose
energy and  angular momentum to the  dark matter particles of  the host. Thus,
dynamical evolution after  accretion can change the properties  of the subhalo
population. A great deal of work has been done to understand the properties of
the   `evolved'  subhalo  population,   using  either   numerical  simulations
\citep[e.g.][]{Klypin99,  Moore99, Sp01b, DMS04,  DKM07, Gao04,  Kang05, Sp08,
  Wetzel10},  or  merger trees  constructed  from  the  extended PS  formalism
\citep[e.g.][]{TB01, TB04, TB05, vdB05, Zen05, Gioc08a, Gan10}.

After accretion, whether a subhalo  can survive as a self-bound entity depends
on its  mass (relative  to that of  its host),  its density profile  (which is
related to  its accretion redshift), and  its orbit. The  model presented here
provides  information   about  the  first  two  parts.   Thus,  combined  with
information    about    the    initial    orbits    of    accreted    subhalos
\citep[e.g.][]{vdB99,  KB06,  Ludlow09,  Valluri10,  Wetzel11} and  about  how
dynamical    friction    and    tidal    stripping   operate    on    subhalos
\citep[e.g.][]{Jiang08, BMQ08}, our model can  be used to construct models for
the  evolved   subhalo  population,  such   as  their  mass   function,  their
distribution   in  host   halos,  and   their  correlation   with   host  halo
properties. We will come back to such modeling in a future paper.

The last  decade has seen much  effort in using halo  occupation statistics to
describe the  galaxy-dark matter  connection in an  attempt to  understand the
galaxy luminosity function, galaxy  clustering, galaxy-galaxy lensing, and the
kinematics  of satellite  galaxies \citep[e.g.][]{Mo99,  BW02,  Yang03, BYM03,
  Zheng05, Tinker05, Cooray05, Cooray06,  CO06, Vale04, Vale06, vdB07, Yang09,
  Cac09,  More09, Moster10}. Several  of these  studies have  tried to  make a
direct link  between the  halo occupation statistics  of galaxies and  of dark
matter  subhalos,  using  abundance  matching  techniques  to  link  satellite
galaxies to subhalos  \citep[e.g.][]{Kravtsov04, Conroy06, Conroy07, Conroy09,
  BCW10,  WJ10,  Wetzel10,  Neistein11,  AvilaReese11}.  These  investigations
typically rely on the un-evolved subhalo mass function, but do not account for
the possibility  that different subhalos  are accreted at different  times and
that the halo mass - galaxy mass relation may be redshift-dependent.  In order
to construct  a self-consistent model  based on abundance matching,  one needs
not only the (un-evolved) subhalo  mass function, but also the distribution of
accretion times. This is exactly what  our model can provide. In a forthcoming
paper  \citep{Yang11},  we will  construct  such  a  self-consistent model  to
characterize how the galaxy-dark matter connection evolves over time.

\section*{Acknowledgements}

We thank Volker  Springel for his help in carrying  out the N-body simulations
through the Partner group collaboration  and the anonymous referee for helpful
comments that  greatly improved the presentation  of this paper.  This work is
supported  by 973 Program  (No.  2007CB815402),  the CAS  Knowledge Innovation
Program (Grant  No.  KJCX2-YW-T05), the CAS  International Partnership Project
``Formation of Galaxies and  Their Activities'' (KJCX2-YW-T23) and grants from
NSFC (Nos. 10821302, 10925314).  HJM  would like to acknowledge the support of
NSF AST-0908334.

\end{document}